\documentclass[prb,preprint,superscriptaddress]{revtex4}
\usepackage{graphicx} 
\usepackage{bm} 
\usepackage{multirow} 
\usepackage{subfigure}
\usepackage{amsfonts}
\usepackage{color}

\begin{document}
 
\title{An analytical law for size effects on thermal conductivity of nanostructures}
 
\author{X. W. Zhou}
\email[]{X. W. Zhou: xzhou@sandia.gov}
\affiliation{Mechanics of Materials Department, Sandia National Laboratories, Livermore, California 94550, USA}

\author{R. E. Jones}
\affiliation{Mechanics of Materials Department, Sandia National Laboratories, Livermore, California 94550, USA}

\date{\today}
 
\begin{abstract}

The thermal conductivity of a nanostructure is sensitive to its dimensions. A simple analytical scaling law that predicts how conductivity changes with the dimensions of the structure, however, has not been developed. The lack of such a law is a hurdle in ``phonon engineering'' of many important applications. Here, we report an analytical scaling law for thermal conductivity of nanostructures as a function of their dimensions. We have verified the law using very large molecular dynamics simulations. 

\end{abstract}
 


\maketitle


Thermal conductivity is a size-independent property for macroscopic scale materials, but becomes sensitive to sizes when the feature dimension is reduced to nano- or micro- meter scale\cite{LWKSYM2003}. This size-dependence is critical for nanostructure applications. For instance, it can be utilized to engineer high thermal conductivity and heat dissipation of microelectronic elements and thereby to effectively increase their density in a device\cite{S2006}. It can also be utilized to engineer low thermal conductivity of thermoelectrics systems to improve energy conversion efficiency\cite{MSS1997}. Previous work\cite{ZAJGS2009,SPK2002,M1992,OS1999,PB1994} has used a Matthiessen rule\cite{LHM2008} to relate thermal conductivity $\kappa$ to sample length $L$:
\begin{equation}
\frac{1}{\kappa\left(L\right)} = \frac{1}{\kappa_b} + \frac{\alpha}{L}
\label{previous kappa}
\end{equation}
where $\kappa_b$ is the thermal conductivity of bulk material and $\alpha$ is a size independent constant. While this rule has been successfully applied\cite{ZAJGS2009,SPK2002}, it is only applicable for heat conduction through the ``thickness'' $L$ of a film (i.e., the sample is assumed to have an infinite cross section). It cannot be applied for heat conduction along a direction in the plane of the film, nor can it be applied for any nanostructures with more than one dimension at the nano- or micro- scale. Because a general equation for thermal conductivity of nanostructures has not been developed, some previous analysis has used the solution of Boltzmann partial differential equations in order to explore the effect of nanostructure sizes\cite{Zhang2007,LLGC2001,LSC2002,WBKR1999,MB2004,VC1999}. This approach is complex and depends on an empirical estimate of the specularity of the free surfaces. It has not been applied to nanostructures with arbitrary dimensions in all three coordinate directions, nor has it provided non-sectioned analytical solutions. The lack of a tractable scaling law has posed a hurdle in ``phonon engineering'' of many nanostructure applications. To overcome this problem, we have developed an analytical scaling law that explicitly expresses thermal conductivity of nanostructures as a function of dimensions in all three coordinate directions. We have also verified the law using large molecular dynamics (MD) simulations.

\begin{figure}[htcp]
\includegraphics[width=3in]{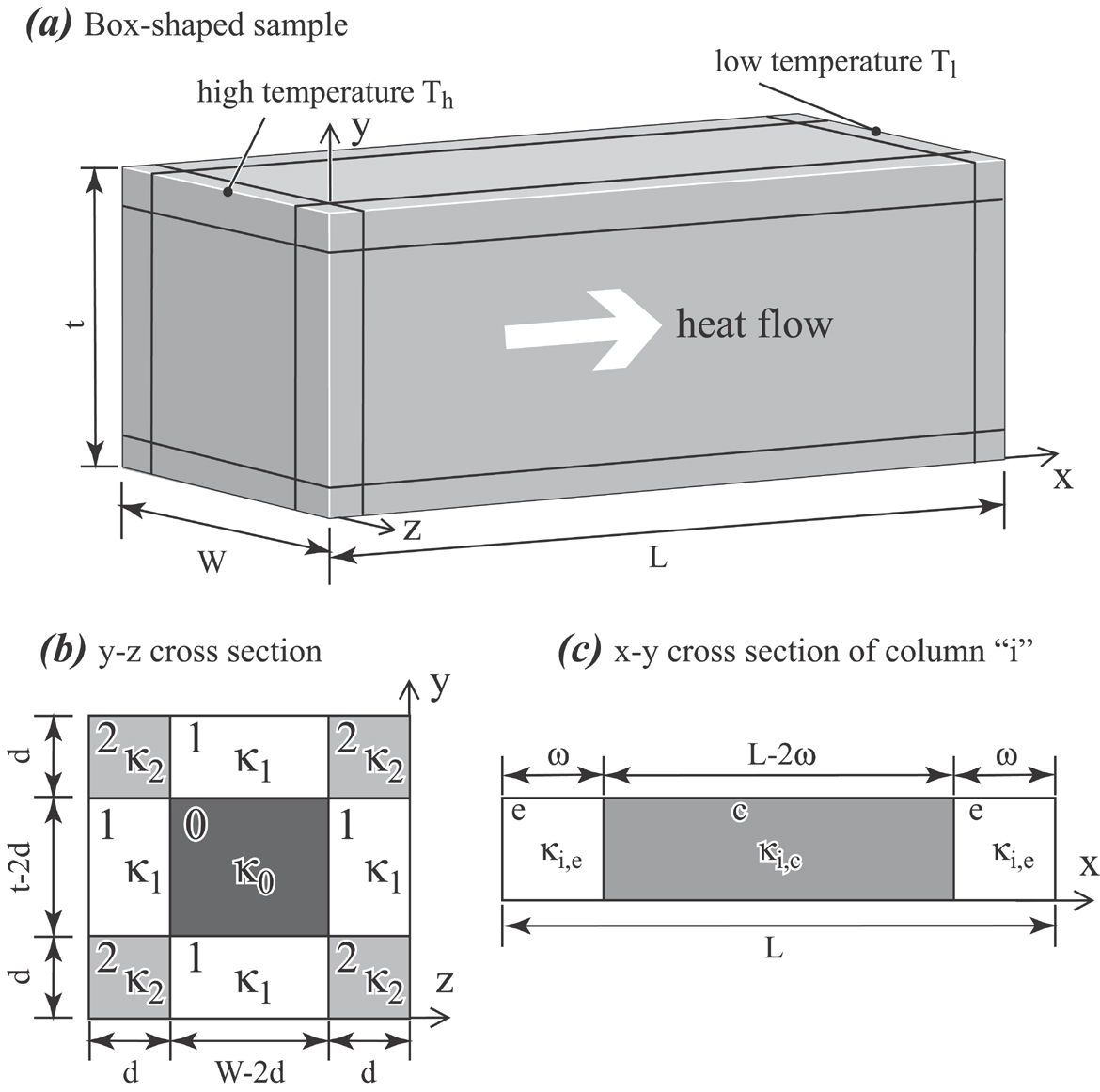}
\caption{Heat conduction through the length $L$ of a box-shaped material with a thickness $t$ and a width $W$.
\label{sample}}
\end{figure}

Consider the unidirectional heat conduction through the length $L$ of a box-shaped sample with a thickness $t$ and a width $W$, as illustrated in Fig. \ref{sample}(a). It is recognized that the size effect on thermal conductivity comes from the surface scattering of phonons, which diminishes as the distance from the surface is increased. Hence, we divide the sample into different regions with respect to the surfaces, Fig. \ref{sample}. First, the (y-z) cross section of the sample is divided into nine regions as shown in Fig. \ref{sample}(b). These nine regions extend in the x- direction into nine small box-shaped pieces (referred to as pillars hereafter) as shown in Fig. \ref{sample}(a). The eight pillars surrounding the center pillar essentially form a shell whose thickness is assumed to be $d$. Each pillar exhibits an apparent mean thermal conductivity throughout length $L$. At a fixed large value of $d$ (say in the order of the phonon mean free path), the boundary environment of each pillar is independent of sample dimension $t$ and $W$. As a result, the apparent thermal conductivity of each pillar is a function of $L$ only. Fig. \ref{sample}(b) indicates that for isotropic materials, the nine pillars fall into three different types $i = 0, 1, 2$, where $i$ refers to the number of the y- or z- surfaces bounding the pillar. Correspondingly, we have three distinguishable conductivity functions $\kappa_0\left(L\right)$, $\kappa_1\left(L\right)$, and $\kappa_2\left(L\right)$. Because heat transports through the pillars in parallel, the overall thermal conductivity of the sample can be calculated as an area-weighted average leading to:
\begin{eqnarray}
&\kappa&\left(t,W,L\right) \nonumber \\ &=& \kappa_0\left(L\right) - \left[\kappa_0\left(L\right) - \kappa_1\left(L\right)\right] \cdot \left(\frac{2d}{t} + \frac{2d}{W}\right) \nonumber \\ &+& \left[\kappa_0\left(L\right) + \kappa_2\left(L\right) - 2\kappa_1\left(L\right)\right] \cdot \frac{4d^2}{t \cdot W}
\label{kappa1}
\end{eqnarray}

Now we consider the thermal transport of each pillar. As shown in Fig. \ref{sample}(c), the pillar can be divided into a center section with a length of $L-2\omega$ and two end sections with a length of $\omega$. At a large given value of $\omega$, the two end surfaces do not interact. As a result, the apparent thermal conductivities of the center and the end sections of the pillar are independent of the length $L$, and therefore can be represented respectively by two constants $\kappa_{i,c}$ and $\kappa_{i,e}$, where subscript $i$ is the pillar number, and $c$ and $e$ designate the center and end sections. Because heat transports through the center and end sections in serial, the overall thermal resistivity (inverse of thermal conductivity) of the pillar can be calculated as the length-weighted average resistivity: $\kappa_{i}^{-1}\left(L\right) = \left(1 - 2\omega/L\right) \cdot \kappa_{i,c}^{-1} + \left(2\omega/L\right) \cdot \kappa_{i,e}^{-1}$, which can be rewritten as
\begin{equation}
\kappa_{i}\left(L\right) = \frac{L \cdot \kappa_{i,c}}{L+\delta_i}
\label{kappa2}
\end{equation}
where $\delta_{i} = 2\omega \cdot \left(\kappa_{i,c}-\kappa_{i,e}\right)/\kappa_{i,e}$ is a positive constant. Substituting Eq. (\ref{kappa2}) into Eq. (\ref{kappa1}), we have an analytical scaling law:
\begin{eqnarray}
&\kappa&\left(t,W,L\right) = \frac{L \cdot \kappa_{0,c}}{L+\delta_0} \nonumber \\ &-& \left[\frac{L \cdot \kappa_{0,c}}{L+\delta_0} - \frac{L \cdot \kappa_{1,c}}{L+\delta_1}\right] \cdot \left(\frac{2d}{t} + \frac{2d}{W}\right) \nonumber \\ &+& \left[\frac{L \cdot \kappa_{0,c}}{L+\delta_0} + \frac{L \cdot \kappa_{2,c}}{L+\delta_2} - 2 \cdot \frac{L \cdot \kappa_{1,c}}{L+\delta_1}\right] \cdot \frac{4d^2}{t \cdot W}
\label{kappa3}
\end{eqnarray}

Eq. (\ref{kappa3}) is valid if a sufficiently large $d$ is used to subsume the surface scattering affected region. Once $d$ is given, Eq. (\ref{kappa3}) involves six parameters $\kappa_{0,c}$, $\kappa_{1,c}$, $\kappa_{2,c}$, $\delta_{0}$, $\delta_{1}$, $\delta_{2}$, where $\kappa_{0,c}$ is essentially the bulk thermal conductivity $\kappa_b$, $\kappa_{1,c}$ is the thermal conductivity near a flat surface, and $\kappa_{2,c}$ is the conductivity near a corner region, see Figs. \ref{sample}(b) and \ref{sample}(c). For MD applications, it is necessary to perform several simulations at different dimensions in order to fit Eq. (\ref{kappa3}). If the minimum dimensions used in these simulations are represented by $t_{min}$ and $W_{min}$, then the largest $d$ that still enables all the MD data to satisfy the geometry condition (i.e., $0 < 2d < t$ and $0 < 2d < W$) is $t_{min}/2$ or $W_{min}/2$ depending on which dimension is smaller.

The model concepts described above can be applied to any sample shapes. Eq. (\ref{kappa3}) also has more general uses. For instance, we found that substituting $t = W = 2r$, $\kappa_{1,c} = \kappa_{2,c}$, and $\delta_{1} = \delta_{2}$ into Eq. (\ref{kappa3}) resulted in the same axial thermal conductivity of a circular wire as a function of wire radius $r$ and length $L$ as we would otherwise derive by directly applying the concepts to the wire case.

When $t \rightarrow \infty$ and $W \rightarrow \infty$, Eq. (\ref{kappa3}) indicates that the inverse of thermal conductivity along the length of sample with an infinite cross section is a linear function of the inverse of the length $1/L$, exactly matching the established equation, Eq. (\ref{previous kappa})\cite{ZAJGS2009,SPK2002,M1992,OS1999,PB1994}. When $L \rightarrow \infty$ and $t \rightarrow \infty$ or $L \rightarrow \infty$ and $W \rightarrow \infty$, Fig. \ref{sample} corresponds to heat transport along a direction in the plane of a large film. Eq. (\ref{kappa3}) then shows that thermal conductivity is a linear function of the inverse of the film thickness ($W$ and $t$ in the two cases both correspond to film thickness). Eq. (\ref{kappa3}) can be effectively verified by checking these linear relationships using either experiments or MD simulations. Here we perform ``direct method'' MD simulations\cite{ZAJGS2009} to verify the linear relationships. The $[0001]$ thermal conductivity of a wurtzite GaN crystal was calculated at a temperature of 300 K. GaN is of interest because many of its applications, such as laser diodes and high electron mobility transistors\cite{JCKSYS2002,ZQWL2003,KCLKRKKC2004,QLGWBL2004,HDCL2002,CJHLKPGSY2003}, operate at high current and power densities where heat dissipation is crucial.

\begin{figure}
\includegraphics[width=2.5in]{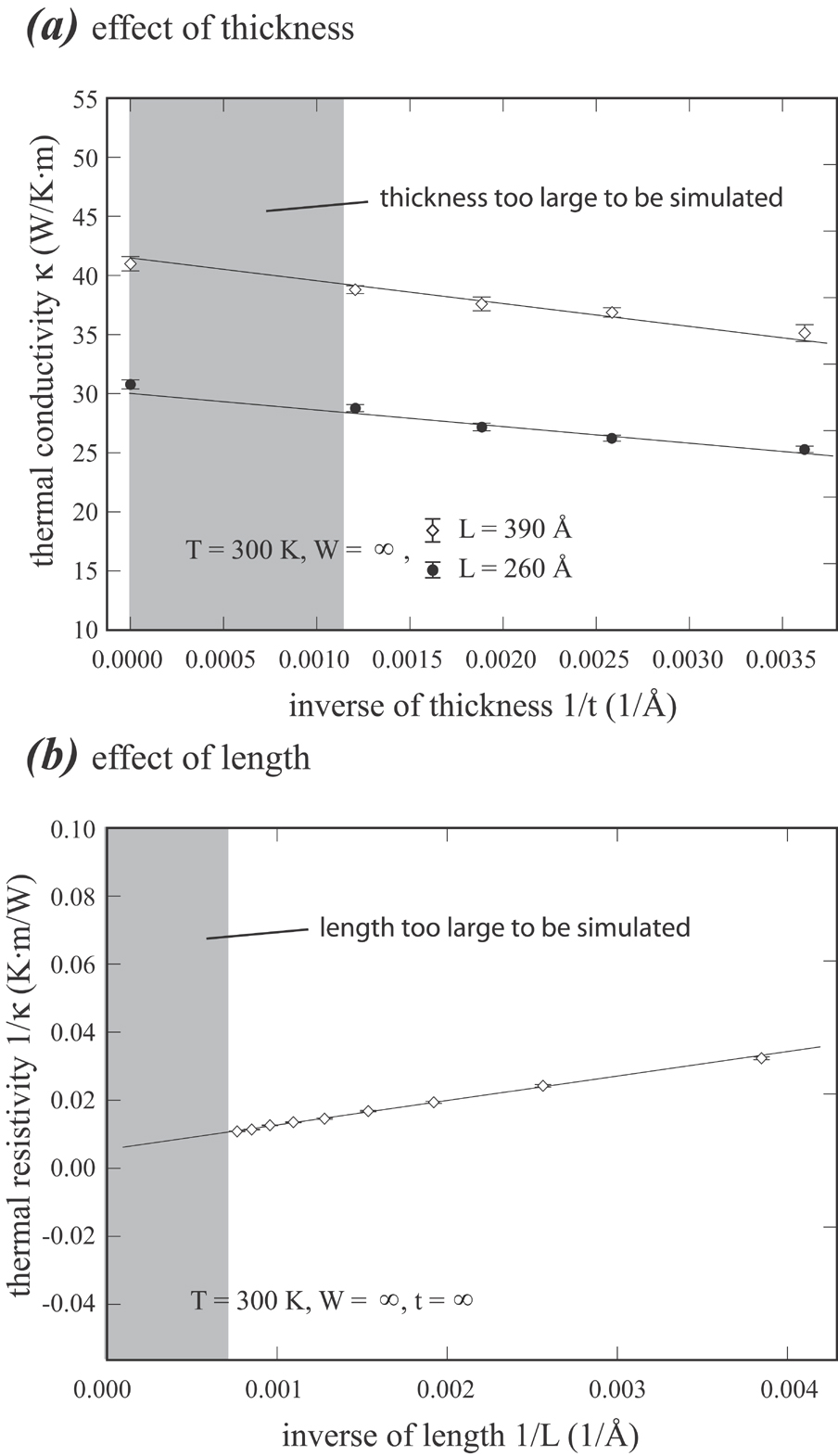}
\caption{GaN thermal conductivity as a function of sample dimension.
\label{300K results}}
\end{figure}

A Stillinger-Weber potential developed by Bere and Serra\cite{BS2002,BS2006} was used. The computational cell is aligned so that the x-, y-, and z- coordinates correspond, respectively, to $[0001]$, $[\bar1100]$, and $[11\bar20]$ directions. A periodic boundary condition was used along the z- axis to simulate an infinite width of $W \rightarrow \infty$, and a free boundary conditions is used in the y- direction to simulate the commonly encountered $[\bar1100]$ surface. Series of thermal conductivities at two lengths (in the x- direction) of $L =$ 260 \AA~and 390 \AA~and different thicknesses (in the y- direction) of $t$ between 276 and 829 \AA~and $t \rightarrow \infty$ (periodic boundary conditions) were calculated. All the simulations applied a very long averaging time of at least 11 ns (some reached 21 ns). Our systems are also relatively large (up to 900,000 atoms). Both a long averaging time and a large number of atoms available for averaging help generate highly accurate results\cite{ZAJGS2009} that strengthened the conclusions. The calculated values of $\kappa$ are shown in Fig. \ref{300K results}(a) against $1/t$, where the lines are produced using Eq. (\ref{kappa3}) with the assigned value $d = 138.13~\AA$ and the fitted parameters $\kappa_{0,c} = 178.38~W/K\cdot m$, $\kappa_{1,c} = 151.65~W/K\cdot m$, $\delta_0 = 1288.10~\AA$, and $\delta_1 = 1329.65~\AA$. It can be seen that the MD data well satisfied the predicted linear relationship and the agreement between the MD data and Eq. (\ref{kappa3}) is excellent. Previously calculated thermal conductivities at different sample lengths $L$ but a fixed sample width $W \rightarrow \infty$ and a fixed sample thickness $t \rightarrow \infty$\cite{ZAJGS2009} are reproduced in Fig. \ref{300K results}(b) using the $1/\kappa$ vs. $1/L$ scale, along with the line generated using Eq. (\ref{kappa3}) with the same parameters. Again the predicted linear relationship is well satisfied and excellent agreement is obtained with only one set of parameters ($d$, $\kappa_{0,c}$, $\kappa_{1,c}$, $\delta_0$, and $\delta_1$) for both thickness and length functions.   

In summary, we have developed an analytical law for size effects on thermal conductivity of nanostructures. This law is very well verified by MD simulations. We expect that it will enable fundamental methods such as MD simulations to be used to study thermal transport at realistic length scales, which would be otherwise impossible due to the limitation of the length scales that can be directly simulated. We also expect that this law can guide experiments to design nanostructured thermal devices. Note that when experimental thermal conductivity data is obtained at different dimensions, the same approach can be used to fit Eq. (\ref{kappa3}). Since experiments are likely to be performed at larger sample dimensions, larger values of $d$ can be chosen to produce even more accurate results. Nonetheless, simulations performed here strongly indicated that even $d = 138.13~\AA$ is sufficient for GaN. 

\begin{acknowledgments}
 
Sandia is a multi-program laboratory operated by Sandia Corporation, a Lockheed Martin Company, for the United States Department of Energy National Nuclear Security Administration under contract DEAC04-94AL85000.This work is performed under a laboratory directed research and development (LDRD) project.

\end{acknowledgments}
 
\appendix
 

\end{document}